# Gauge Invariant States of QCD

Scott Chapman

*Chapman University, One University Drive, Orange, CA 92866*
(Dated: February 29, 2012)

Abstract

None of the asymptotic states commonly used in perturbative QCD are gauge invariant. A similar statement could be made about QED, but in QED one can construct gauge invariant "dressed" states (with Dirac electrons) that are unitarily equivalent to the states used in perturbation theory. Perturbative expansions of dressed states have been derived for QCD, but Gribov copies obstruct these states from remaining gauge invariant non-perturbatively. Introduced here are new QCD states that are exact solutions to the Dirac condition, so they remain gauge invariant non-perturbatively. The quark content of these states is restricted to be meson-like, baryon-like, or anti-baryon-like locally at each point of space. The states differ from others previously presented due to the use of creation operators defined in coordinate space rather than in momentum space.

## **Introduction**

Both the QED and QCD gauge symmetries are thought to be unbroken in Nature. For this to be true, every allowed physical state must be invariant to QED and QCD gauge transformations. However, none of the asymptotic states commonly used in perturbation theory meet this criteria.

In QED this is not a problem, since it is possible to construct a vacuum and a complete set of gauge invariant states by "dressing" the electron operators (and other fermion fields) with exponents involving longitudinal photon fields [1,2]. States made from these dressed operators are unitarily equivalent to the states normally used in QED perturbation theory.

When moving to QCD, things become more complicated. Quark and gluon operators can be dressed perturbatively [3-9], but Gribov copies obstruct these operators from being gauge invariant non-perturbatively [5-9]. One can construct colorless states from dressed quarks and gluons, but the dressing itself is still defined perturbatively.

The purpose of this paper is to present a set of exactly gauge invariant QCD states – ones that do not rely on perturbative expansions to achieve gauge invariance. The paper begins with a review of Dirac's method of consistently separating states into "physical" states that are gauge invariant and "unphysical" states that are not. Using this method, gauge invariant QED states are constructed in a new way that is directly generalizable to QCD. It is shown that this new approach arrives at the same results (Dirac electrons) as the standard approach used to create gauge invariant QED states.

The new approach is applied to QCD in order to derive an exactly gauge invariant "physical" zero state. Gauge invariant local operators are then derived which can act on the zero state to create other "physical" states. One finds that gauge invariance restricts the quark content of these states to be meson-like, baryon-like, or anti-baryon-like; single-quark states are not gauge invariant. The most interesting feature of these states is the fact that the creation operators used to build them are defined at each point in coordinate space, rather than at each point in momentum space like standard creation operators. Since the Dirac condition is local, gauge invariance demands colorless quark combinations locally at each point of space, not just when integrated over all space.

## **Dirac Quantization**

In every gauge theory, there is a local constraint operator $G^a(x)$ that commutes with the Hamiltonian. The constraint operator for a given gauge theory is most easily found by deriving the Hamiltonian from the Lagrangian. The Hamiltonian derived this way includes a term proportional to $A_0^a$ (the temporal component of the gauge field). The coefficient of that term is the constraint operator $G^a(x)$ for that gauge theory [10].

The constraint operator is an integral part of any gauge theory, containing important physics that is not contained in the Hamiltonian. For example, in QED, the Hamiltonian equations alone do not produce Gauss' Law. Rather, Gauss' Law is separately enforced by setting QED's constraint operator equal to zero. Constraint operators also generate gauge transformations through their commutation relations. For example in SU(N),

$$[G^a(t,\vec{x}), G^b(t,\vec{y})] = if^{abc}G^c(t,\vec{x})\delta^3(\vec{x}-\vec{y}), \tag{1}$$

where $f^{abc}$ are the structure functions of the SU(N) group. As a consequence, $|\xi\rangle \to \exp(i\varepsilon^a G^a(x))|\xi\rangle$ is a local gauge transformation on the state $|\xi\rangle$.

In order for a state to be gauge invariant, it should not be affected by this kind of gauge transformation. There are two standard ways to achieve gauge invariance for states: Reduction or Dirac quantization [1,10,11].

a) Reduction: Set $G^a(x) = 0$ (operator equation) (2)

b) Dirac Quantization: A state $|\xi\rangle$ is only "physical" if $G^a(x)|\xi\rangle = 0$ (3)

The Reduction approach can be used exactly for Abelian theories like QED or perturbatively for non-Abelian theories. In QED, it amounts to enforcing Gauss' Law at an operator level – equating the divergence of the electric field operator to the charge density of matter field operators. Reduction also requires a gauge condition in order to fully remove the longitudinal gauge field degree of freedom. The Reduction approach is spelled out explicitly in some of the older QED texts (see for example [12]).

Dirac Quantization is an alternative approach to enforcing gauge invariance that arrives at the same physical results. The restriction that any physical state must be annihilated by the constraints is consistent since gauge constraints commute with the Hamiltonian. When Dirac Quantization is used for QED, longitudinal photon operators are retained in the theory, but states involving them are considered unphysical since they do not obey the Dirac condition (3). Dirac quantization of QED also requires use of a gauge-invariant Dirac electron [1,2] operator that differs from but is unitarily equivalent to the standard electron operator. For non-Abelian gauge theories like QCD, Dirac Quantization is the only option for maintaining exact gauge invariance, since the Reduction approach is only solvable perturbatively.

To portray the main points of Dirac quantization, QED is addressed first, and a complete set of gauge-invariant QED states is derived. These states are built from the usual transverse photons and Dirac electrons, but a new method of derivation is employed that can be generalized to produce exactly gauge invariant QCD states. QCD is addressed next, and the results mentioned above are derived.

**Dirac Quantization of QED**

The QED Hamiltonian is given by

$$H = \int d^3x \{\tfrac{1}{2}\Pi_i \Pi_i + \tfrac{1}{4} F_{ij} F_{ij} - \overline{\psi}[\gamma^i(i\partial_i - eA_i) - m]\psi\}, \quad (4)$$

where $\Pi_i$ are electric field operators, $F_{ij} \equiv (\partial_i A_j - \partial_j A_i)$ are magnetic field operators, and $\psi$ are the matter fields. The quantized fields and their canonical momenta obey the following equal-time commutation relations:

$$[\Pi_i(t,\vec{x}), A_j(t,\vec{y})] = -i\delta_{ij}\delta^3(\vec{x}-\vec{y})$$

$$\{\psi(t,\vec{x}), \psi^\dagger(t,\vec{y})\} = \delta^3(\vec{x}-\vec{y}).  \qquad (5)$$

Like all gauge theories, QED has a local constraint operator that commutes with the Hamiltonian:

$$G(x) \equiv \partial_i \Pi_i(x) - e\psi^\dagger(x)\psi(x) . \qquad (6)$$

From the form of the above equation (and keeping in mind that $\Pi_i$ is the electric field), one can see that setting $G(x)=0$ everywhere imposes the local form of Gauss' Law at an operator level (Reduction approach). As explained in the Introduction, setting $G(x)=0$ also ensures that quantum states are gauge invariant.

Alternatively, Gauss' Law and gauge invariance can be enforced through Dirac Quantization as expressed in equation (3). The Dirac condition is local since it applies to every point of spacetime, and it is a consistent separation of physical from unphysical states since $[G(x), H] = 0$. The advantage to Dirac's approach (especially for non-Abelian theories) is that it does not require one to keep track of the complicated field substitutions that can arise when enforcing Gauss' Law at an operator level. Equation (3) ensures that the contribution of the constraint (6) will be zero inside any matrix element involving physical states.

Throughout this paper, when only a spatial dependence is shown (e.g. $\psi(\vec{x})$), it will apply to a field at time $t=0$. When a spacetime dependence is shown (e.g. $G(x)$), it will apply to an operator that commutes with the Hamiltonian, meaning that its form at all times is the same as that at $t=0$.

**Pure-gauge QED in the absence of matter fields**

In order to make this QED analysis more closely parallel to the QCD analysis of the next section, Dirac quantization of pure-gauge QED in the absence of matter fields will be considered first. One may start with the canonical momentum state $|\Pi_0\rangle$ defined at each point of space by:

$$\Pi_i(\vec{x})|\Pi_0\rangle = 0. \qquad (7)$$

In the absence of matter fields, the state $|\Pi_0\rangle$ clearly satisfies the Dirac condition $G(x)|\Pi_0\rangle = \partial_i \Pi_i(\vec{x})|\Pi_0\rangle = 0$. However, the state $|\Pi_0\rangle$ is not an acceptable physical state due to the fact that it has an ill-defined norm. This is because in order to calculate the norm of the momentum-representation state $|\Pi_0\rangle$, one must insert a complete set of coordinate-representation states $|A_i(\vec{x})\rangle\langle A_i(\vec{x})|$ and functionally integrate over all possible values of $A_i(\vec{x})$ at each point of space. The resulting integrals are infinite, so the norm is infinite. Below it will be shown that the definition of the scalar product should include an operator insertion that eliminates the infinities in the pure-gauge (longitudinal) direction. But this is not enough, because $\langle\Pi_0\|\Pi_0\rangle$ also has infinite integrals in non-gauge (transverse) directions.

One way to regulate these transverse integrals is to introduce a Gaussian factor in front of $|\Pi_0\rangle$ for each point of space. For example, denoting the transverse components of $A_i(\vec{x})$ by $A_{Ti}(\vec{x})$, the norm of the state $\prod_x \exp\left(-\tfrac{1}{2}\mu^{-2} A_{Ti}(\vec{x}) A_{Ti}(\vec{x})\right)|\Pi_0\rangle$ has finite integrals in the transverse directions. Since the transverse directions are defined via derivatives, a cleaner way to regulate the transverse directions is to use $|\tilde{0}_\Lambda\rangle \equiv \exp\left(-\tfrac{1}{4}\Lambda^{-1}\int d^3x F_{ij} F_{ij}\right)|\Pi_0\rangle$, where $\Lambda$ is some mass scale introduced to make the exponent dimensionless, and the product of Gaussian factors at each point of space has been converted to a sum (or integral) in the exponent. The state $|\tilde{0}_\Lambda\rangle$ satisfies the Dirac condition (since $[G(x'), F_{ij}(\vec{x})] = 0$), and the only infinities in $\langle\tilde{0}_\Lambda\|\tilde{0}_\Lambda\rangle$ occur in the longitudinal integrals (which will be regulated). So $|\tilde{0}_\Lambda\rangle$ is an acceptable state upon which to build a complete set of physical states for QED. However, a disadvantage to using this state is that it involves an arbitrary mass scale that was inserted by hand.

For QED, there is another option. Since the gauge constraint is Abelian, the following dimensionless operator is gauge invariant

$$O_T \equiv -\tfrac{1}{4}\int \frac{d^3z\, d^3y\, d^3k}{(2\pi)^3 k} e^{i\vec{k}\cdot(\vec{z}-\vec{y})} F_{ij}(\vec{z}) F_{ij}(\vec{y}). \tag{8}$$

In other words, $[G(x), O_T] = 0$. Therefore the state

$$|\tilde{0}\rangle \equiv \exp(O_T)|\Pi_0\rangle \tag{9}$$

satisfies the Dirac condition. The factor $\exp(O_T)$ also brings Gaussian dependencies in transverse directions that regulate the integrals in those directions when calculating the state's norm. The advantage of the state $|\tilde{0}\rangle$ over the state $|\tilde{0}_\Lambda\rangle$ is that the former regulates transverse integrals without the introduction of an arbitrary mass scale. Another way to see that $|\tilde{0}\rangle$ is regular in its transverse directions is as follows: First expand the gauge fields into the usual plane waves:

$$A_i(\vec{x}) = \int \frac{d^3k}{\sqrt{2k(2\pi)^3}} \varepsilon_i^\lambda(\vec{k}) \left( a_\lambda(\vec{k}) e^{i\vec{k}\cdot\vec{x}} + a_\lambda^\dagger(\vec{k}) e^{-i\vec{k}\cdot\vec{x}} \right)$$

$$\Pi_i(\vec{x}) = -i \int \frac{d^3k \sqrt{k}}{\sqrt{2(2\pi)^3}} \varepsilon_i^\lambda(\vec{k}) \left( a_\lambda(\vec{k}) e^{i\vec{k}\cdot\vec{x}} - a_\lambda^\dagger(\vec{k}) e^{-i\vec{k}\cdot\vec{x}} \right). \quad (10)$$

Using these momentum expansions, it is possible to see that $a_{T\lambda}(\vec{k})|\tilde{0}\rangle = 0$, where $a_{T\lambda}(\vec{k})$ are the transverse photon destruction operators. This means that in the transverse photon directions, $|\tilde{0}\rangle$ is equivalent to the normal perturbative vacuum of pure-gauge QED. That vacuum state has finite norm; and therefore $|\tilde{0}\rangle$ also has a finite norm in the transverse directions.

When using the Dirac condition, the scalar product always has to be regulated. This is because the naïve inner product of any two states that obey the Dirac condition (3) always has infinite integrals in its pure gauge directions. This problem is unavoidable and expected, since the Dirac condition only constrains the pure-gauge canonical momentum ($\partial_i \Pi_i(\vec{x})$ for QED) without also constraining its pure-gauge field counterparts ($\partial_i A_i(\vec{x})$ for QED). The standard way to address this issue is to include a delta function and a determinant in the definition of a physical scalar product. Specifically, the scalar product for QED in the Dirac approach is defined as follows [10]:

$$(\xi'|\xi) \equiv \langle \xi' | \det[\chi, G] \delta(\chi) | \xi \rangle \qquad \chi(\vec{x}) = \partial_i A_i(\vec{x}). \quad (11)$$

The function $\chi(\vec{x})$ can actually be any function that satisfies $\det[\chi, G] \neq 0$, but it is often convenient to choose $\chi(\vec{x}) = \partial_i A_i(\vec{x})$. It is apparent that with this choice of $\chi(\vec{x})$, the delta function does indeed remove the infinite longitudinal integrals at each point of space. As an aside it should be noted that the BRST formalism arrives at the same result without an operator insertion by expanding the functional space to include canonically quantized ghost fields. Integration over ghosts fields yields the Fadeev-Popov determinant $\det[\chi, G]$, while integration over a non-minimal canonical momentum field yields $\delta(\chi)$ [10,13,14].

In the absence of matter fields, the state $|\tilde{0}\rangle$ is a gauge invariant physical state that satisfies that Dirac condition and has finite norm. Moreover since $[G(x), a_{T\lambda}^\dagger(\vec{k})] = 0$, any state made by adding transverse creation operators $a_{T\lambda}^\dagger(\vec{k})$ to $|\tilde{0}\rangle$ is also gauge invariant and physical.

**QED with matter fields**

To add matter fields to the Dirac quantization of QED, it is helpful to first make the usual plane-wave expansions:

$$\psi(\vec{x}) = \int \frac{d^3 p}{\sqrt{2E_p (2\pi)^3}} \{e^{i\vec{p}\cdot\vec{x}} u_s(\vec{p}) b_s(\vec{p}) + e^{-i\vec{p}\cdot\vec{x}} v_s(\vec{p}) d_s^\dagger(\vec{p})\} . \tag{12}$$

One may then expand the definition of $|\tilde{0}\rangle$ to include matter fields by requiring

$$b_s(\vec{p})|\tilde{0}\rangle = d_s(\vec{p})|\tilde{0}\rangle = 0 . \tag{13}$$

Through that definition, $|\tilde{0}\rangle$ becomes equivalent to the normal perturbative vacuum of QED (except in the longitudinal photon direction). Using (12), it is apparent that the full constraint $G(x) \equiv \partial_i \Pi_i(x) - e\psi^\dagger(x)\psi(x)$ has terms in it proportional to $b_{s'}^\dagger(\vec{p}')d_s^\dagger(\vec{p})$, involving two creation operators. As a result, $G(x)|\tilde{0}\rangle \neq 0$, the Dirac condition is not satisfied, and $|\tilde{0}\rangle$ is no longer gauge invariant once quarks are included.

However, a state which does satisfy the Dirac condition can easily be constructed. Consider the Hermitian operator

$$O_L \equiv e \int \frac{d^3x\, d^3y\, d^3k}{(2\pi)^3 k^2} e^{i\vec{k}\cdot(\vec{x}-\vec{y})} \partial_i A_i(\vec{x}) \psi^\dagger(\vec{y}) \psi(\vec{y}) . \tag{14}$$

The field commutation relations can be used to derive the following operator equation:

$$G(x)\exp(iO_L) = \exp(iO_L)\partial_i \Pi_i(\vec{x}) . \tag{15}$$

In other words, when $G(x)$ is commuted through $\exp(iO_L)$ the matter term $-e\psi^\dagger(\vec{x})\psi(\vec{x})$ is removed from the constraint $G(x)$. Therefore, the state

$$|0\rangle \equiv \exp(iO_L)|\tilde{0}\rangle \tag{16}$$

is gauge invariant, satisfying the Dirac condition $G(x)|0\rangle = 0$.

Once a gauge-invariant perturbative vacuum state $|0\rangle$ with finite norm has been established, it is possible to define additional gauge-invariant states through the application of gauge-invariant creation operators. Defining dressed operators by

$$\tilde{b}_s^\dagger(\vec{p}) \equiv \exp(iO_L) b_s^\dagger(\vec{p}) \exp(-iO_L)$$
$$\tilde{d}_s^\dagger(\vec{p}) \equiv \exp(iO_L) d_s^\dagger(\vec{p}) \exp(-iO_L) , \tag{17}$$

one can see from (15) that

$$\left[G(x), \tilde{b}_s^\dagger(\vec{p})\right] = \left[G(x), \tilde{d}_s^\dagger(\vec{p})\right] = 0 , \tag{18}$$

so they are gauge invariant. The operators $\tilde{b}_s^\dagger(\vec{p})$ and $\tilde{d}_s^\dagger(\vec{p})$ are just the Dirac electron "dressed" operators that have long been known to be gauge invariant [1,2]. Any state formed by applying these operators or transverse photon creation operators $a_{T\lambda}^\dagger(\vec{k})$ to the state $|0\rangle$ is a gauge-invariant physical state.

**Dirac Quantization of QCD**

Many of the steps taken above for QED can be repeated in order to create gauge-invariant states for QCD. The QCD Hamiltonian and its corresponding constraints are:

$$H = \int d^3x \left\{ \tfrac{1}{2}\Pi_i^a \Pi_i^a + \tfrac{1}{4} F_{ij}^a F_{ij}^a - \overline{\psi}(\gamma^i iD_i - m)\psi \right\} \tag{19}$$

$$G^a(x) \equiv D_i^{ab}\Pi_i^b + g\psi^\dagger \lambda^a \psi , \tag{20}$$

where $F_{ij}^a \equiv \partial_i A_j^a - \partial_j A_i^a + gf^{abc}A_i^b A_j^c$, $D_i \equiv \partial_i - igA_i^a \lambda^a$, $D_i^{ab} \equiv \delta^{ab}\partial_i - gf^{abc}A_i^c$ and $\lambda^a$ are half the Gell-Mann matrices, satisfying $[\lambda^a, \lambda^b] = if^{abc}\lambda^c$. As in QED, the local constraints $G^a(x)$ commute with the Hamiltonian and generate gauge transformations. The operator equations $G^a(x) = 0$ are the color equivalent of Gauss' Law, which should be imposed one way or another to preserve gauge invariance. Solving the nonlinear operator equation $G^a(x)=0$ is only possible perturbatively, but the Dirac condition (3) can be enforced exactly.

**Pure-gauge QCD in the absence of matter fields**

Proceeding in parallel to the QED derivation and starting with pure-gauge QCD, it is apparent that the canonical momentum state $|\Pi_0\rangle$ defined by

$$\Pi_i^a(\vec{x})|\Pi_0\rangle = 0 \tag{21}$$

satisfies $D_i^{ab}\Pi_i^b|\Pi_0\rangle = 0$ and is therefore invariant to gauge transformations in the absence of matter fields. But just as in the QED case, $|\Pi_0\rangle$ has an ill-defined norm due to infinite functional integrals over $A_i(\vec{x})$ field configurations. In QED, the norm in transverse directions can be made finite by introducing a gauge invariant operator $\exp(O_T)$ (eqn. 8) that essentially transforms $|\Pi_0\rangle$ into the perturbative vacuum. However, if one generalizes the QED operator $O_T$ to QCD, one finds that the non-Abelian version does not commute with the constraints, so it is not gauge invariant:

$$\left[G^a(x), \int \frac{d^3z\, d^3y\, d^3k}{(2\pi)^3 k} e^{i\vec{k}\cdot(\vec{z}-\vec{y})} F_{ij}^b(\vec{z}) F_{ij}^b(\vec{y})\right] \neq 0. \tag{22}$$

However, the commutator (22) is proportional to the coupling constant $g$. As a result, it is possible to create a perturbative expression for the QCD analog of $\exp(O_T)$ in QED. This exponent can then be used to create QCD dressed states that are perturbatively gauge invariant [3-9]. But as shown in [5-9], Gribov copies ruin the full gauge invariance of these dressed states. In this paper, a different operator will be introduced which will lead to finite-norm states that exactly commute with the gauge constraints and are therefore exactly gauge invariant.

The operator $\exp(O_T)$ for QED had an interesting property that was not explicitly discussed in that section. Namely, the state $\exp(O_T)|\Pi_0\rangle$ is an exact eigenstate of the pure-gauge QED Hamiltonian. So if one could find an operator that would create an exact eigenstate of the pure-gauge QCD Hamiltonian, perhaps that operator could also be used to regulate non-gauge integrals in norms. Consider the operator:

$$O_0 \equiv \tfrac{1}{2}\varepsilon_{ijk} \int d^3x \left(\partial_i A_j^a A_k^a + \tfrac{1}{3} g f^{abc} A_i^a A_j^b A_k^c\right) \tag{23}$$

This operator is gauge-invariant

$$[G^a(x), O_0] = 0, \qquad (24)$$

and its exponent can remove the magnetic energy term from the pure-gauge Hamiltonian:

$$\int d^3x \{\tfrac{1}{2}\Pi_i^a\Pi_i^a + \tfrac{1}{4}F_{ij}^a F_{ij}^a\}\exp(O_0) = \exp(O_0)\int d^3x (\tfrac{1}{2}\Pi_i^a\Pi_i^a - \tfrac{1}{2}i\varepsilon_{ijk}F_{ij}^a\Pi_k^a). \qquad (25)$$

Because of the above equations, $\exp(O_0)|\Pi_0\rangle$ is exactly gauge invariant and is also an exact eigenstate of the pure-gauge QCD Hamiltonian with vanishing eigenvalue. Unfortunately when the norm of $\exp(O_0)|\Pi_0\rangle$ is calculated, one of the two transverse integrals of $A_i^a(\vec{x})$ explodes at each point of space, so $\exp(O_0)|\Pi_0\rangle$ does not have a finite norm.

Just as in QED, another way to maintain exact gauge invariance of states while regulating transverse integrals over $A_i^a(\vec{x})$ in their scalar products is to use an operator involving an arbitrary mass scale. Consider the operator

$$O_\Lambda \equiv -\tfrac{1}{4}\Lambda^{-1}\int d^3x F_{ij}^a F_{ij}^a, \qquad (26)$$

which is made dimensionless by insertion of a mass scale $\Lambda$. The operator is gauge-invariant, $[G^a(x), O_\Lambda] = 0$, so the state

$$|\tilde{0}_\Lambda\rangle \equiv \exp(O_\Lambda)|\Pi_0\rangle \qquad (27)$$

is exactly gauge invariant (for pure-gauge transformations). Thanks to the Gaussian exponent, when the scalar product prescription of (11) is used, $|\tilde{0}_\Lambda\rangle$ has finite norm.

One may ask what is the meaning of the mass scale $\Lambda$ introduced here and whether it has any relation to $\Lambda_{QCD}$. To answer that question, it should be noted that $|\tilde{0}_\Lambda\rangle$ has a zero-point energy whose dominant contribution is proportional to $\int d^3x d^3k (k^2/\Lambda)$. This energy is minimized by the mass scale $\Lambda$ becoming as large as possible (e.g. the Planck scale). So it would seem that $\Lambda$ is more akin to a regulator of divergences than it is to $\Lambda_{QCD}$. Answering the question more fully would require developing a perturbation theory from these gauge invariant states which is outside the scope of this paper.

Once $|\tilde{0}_\Lambda\rangle$ has been defined, other pure-gauge-invariant states can be formed by applying other pure-gauge-invariant operators. To do this, it is helpful to define the following pure-gauge-invariant local operators:

$$\mathcal{H}_E(\vec{x}) = \tfrac{1}{2}\Pi_i^a(\vec{x})\Pi_i^a(\vec{x}) \qquad \mathcal{H}_M(\vec{x}) = \tfrac{1}{4}F_{ij}^a(\vec{x})F_{ij}^a(\vec{x}). \tag{28}$$

Both of these Hamiltonian density operators commute with the constraint $G^a(x)$. Therefore, one can define a set of physical states for pure-gauge QCD as the state $|\tilde{0}_\Lambda\rangle$ along with any combination of the operators $\mathcal{H}_E(\vec{x})$, $\mathcal{H}_M(\vec{x})$ and $O_0$ (or their commutators) acting on $|\tilde{0}_\Lambda\rangle$. That definition of physical states may not be the complete set of states that satisfy the Dirac condition for pure-gauge QCD, but it is a consistent closed set in the sense that the Hamiltonian acting on any state in the set results in another state in the set.

**Dirac Quantization of QCD with Quarks**

Building on the results of the last section, one can now add quarks to the theory. To better highlight the chiral aspects of quarks, a massless expansion of quark fields will be used in this section, even though quark masses are included in the Hamiltonian (19). Employing the Weyl representation of the gamma matrices

$$\gamma^0 = -\begin{pmatrix} 0 & 1 \\ 1 & 0 \end{pmatrix} \qquad \gamma^i = \begin{pmatrix} 0 & \sigma^i \\ -\sigma^i & 0 \end{pmatrix} \qquad \gamma_5 = \begin{pmatrix} 1 & 0 \\ 0 & -1 \end{pmatrix}, \tag{29}$$

the quark fields can be expanded as follows at time $t = 0$:

$$\psi(\vec{x}) = b(\vec{x}) + d^\dagger(\vec{x}) = \int \frac{d^3 p}{2p^0 \sqrt{(2\pi)^3}} \left\{ \begin{pmatrix} p\cdot\bar{\sigma} S_s \\ p\cdot\sigma S_s \end{pmatrix} b_s(\vec{p}) + \begin{pmatrix} p\cdot\sigma S_s \\ p\cdot\bar{\sigma} S_s \end{pmatrix} d_s^\dagger(-\vec{p}) \right\} e^{i\vec{p}\cdot\vec{x}}, \tag{30}$$

where $p^0 = |\vec{p}|$, $\sigma^\mu \equiv (1, \sigma^i)$, $\bar{\sigma}^\mu \equiv (1, -\sigma^i)$, and $S_s$ is a 2-component spin unit vector. The local fields $b(\vec{x})$ and $d^\dagger(\vec{x})$ are defined as the parts of $\psi(\vec{x})$ that involve all of the quark destruction operators $b_s(\vec{p})$ or all of the anti-quark creation operators $d_s^\dagger(-\vec{p})$, respectively. Just as in QED, one finds that the quark part of the constraint $G^a(x)$ (defined in (20)) has terms in it proportional to two quark creation operators. As a result, $G^a(x)$ does not annihilate the standard perturbative QCD vacuum, meaning it is not gauge-invariant. Similarly, enlarging the definition of $|\Pi_0\rangle$ in (21) to require $b_s(\vec{p})|\Pi_0\rangle = d_s(\vec{p})|\Pi_0\rangle = 0$, one finds that $G^a(x)|\tilde{0}_\Lambda\rangle \neq 0$. In other words, although $|\tilde{0}_\Lambda\rangle$ is invariant to pure-gauge transformations, it is no longer gauge-invariant when quarks are included.

In QED, to get rid of terms like $b_{s'}^\dagger(\vec{p}')d_s^\dagger(\vec{p})$ in $G(x)$, it is possible to construct an exponent $\exp(iO_L)$ involving the exact canonical conjugate of the photon part of the constraint, and to use that to completely cancel the matter part of the constraint. A perturbative version of that method can be used for QCD [4-9], but a version that remains valid non-perturbatively is not available since one cannot write down a non-perturbative analytical expression for the canonical conjugate of $D_i^{ab}\Pi_i^b$. However for QCD, it is possible to construct an exponent involving only quarks that still enables construction of exactly gauge-invariant states with certain, more restricted configurations.

Consider the following Hermitian operator

$$O_\psi \equiv \tfrac{1}{4}\pi \int \frac{d^3p}{E} S_{s'}^T p \cdot \vec{\sigma} S_s \left(b_{s'}^\dagger(\vec{p})d_s^\dagger(-\vec{p}) + d_{s'}(-\vec{p})b_s(\vec{p})\right)$$
$$= -\tfrac{1}{4}\pi \int d^3x \left(b^\dagger(\vec{x})\gamma^0(1-\gamma_5)d^\dagger(\vec{x}) + d(\vec{x})\gamma^0(1+\gamma_5)b(\vec{x})\right), \qquad (31)$$

where the second line simply rewrites the first line using the spatial operators defined in (30). The following commutation relations hold:

$$[b(\vec{x}), O_\psi] = -\tfrac{1}{4}\pi\gamma^0(1-\gamma_5)d^\dagger(\vec{x})$$
$$[d^\dagger(\vec{x}), O_\psi] = -\tfrac{1}{4}\pi\gamma^0(1+\gamma_5)b(\vec{x}). \qquad (32)$$

Using these relations together with the Hadamard Lemma of the Baker Hausdorff theorem [15], one finds the following relation for the quark part of the constraint:

$$\psi^\dagger(\vec{x})\lambda^a\psi(\vec{x})\exp(iO_\psi) = \exp(iO_\psi)\left[\tilde{b}^\dagger(\vec{x})\lambda^a\tilde{b}(\vec{x}) + \tilde{d}(\vec{x})\lambda^a\tilde{d}^\dagger(\vec{x})\right], \qquad (33)$$

where the local operators $\tilde{b}(\vec{x})$ and $\tilde{d}^\dagger(\vec{x})$ are defined by

$$\tilde{b}(\vec{x}) \equiv \frac{1}{\sqrt{2}}(1-i\gamma^0\gamma_5)b(\vec{x}) \qquad \tilde{d}^\dagger(\vec{x}) \equiv \frac{1}{\sqrt{2}}(1+i\gamma^0\gamma_5)d^\dagger(\vec{x}). \qquad (34)$$

In other words, commuting $\psi^\dagger(\vec{x})\lambda^a\psi(\vec{x})$ through the factor $\exp(iO_\psi)$ separates the quark part of the constraint from the anti-quark part, and the constraint no longer has terms involving two creation operators.

With these tools in place, one can define an exactly gauge invariant state for QCD that is valid non-perturbatively:

$$|0_\Lambda\rangle \equiv \exp(iO_\psi)|\tilde{0}_\Lambda\rangle . \tag{35}$$

Due to (33) and the fact that $\lambda^a$ is traceless, this state satisfies the Dirac condition with quarks: $G^a(x)|0_\Lambda\rangle = 0$. Moreover, it has a well-defined norm when the scalar product definition of (11) is used.

Once one gauge-invariant state has been defined, it is possible to define others. Although it is not possible to construct single-particle quark or gluon operators that are non-perturbatively gauge invariant [5-9], it is possible to construct multi-quark states that exactly obey the Dirac condition and are therefore exactly gauge invariant (not just perturbatively). This task is greatly simplified by the fact that the operators defined in (34) and (30) have local commutation relations:

$$\{\tilde{b}(\vec{x}),\tilde{b}^\dagger(\vec{y})\} = \delta^3(\vec{x}-\vec{y})\tfrac{1}{2}(1-i\gamma^0\gamma_5)$$

$$\{\tilde{d}(\vec{x}),\tilde{d}^\dagger(\vec{y})\} = \delta^3(\vec{x}-\vec{y})\tfrac{1}{2}(1+i\gamma^0\gamma_5). \tag{36}$$

Denoting quantities "inside" the exponent $\exp(iO_\psi)$ with a tilde, one finds

$$\tilde{G}^a(x) = \exp(-iO_\psi)G^a(x)\exp(iO_\psi) = D_i^{ab}\Pi_i^b + g\tilde{b}^\dagger\lambda^a\tilde{b} + g\tilde{d}\lambda^a\tilde{d}^\dagger \tag{37}$$

$$\tilde{H} = \exp(-iO_\psi)H\exp(iO_\psi) = \int d^3x\left[\mathcal{H}_E(\vec{x}) + \mathcal{H}_M(\vec{x}) + \tilde{\mathcal{H}}_\psi(\vec{x}) + \tilde{\mathcal{H}}_m(\vec{x})\right] \tag{38}$$

with $\tilde{\mathcal{H}}_\psi(\vec{x}) = \tilde{b}^\dagger\gamma^i D_i\tilde{b} + \tilde{d}\gamma^i D_i\tilde{d}^\dagger$ and $\tilde{\mathcal{H}}_m(\vec{x}) = \tilde{b}^\dagger\gamma^0 m\tilde{d}^\dagger + \tilde{d}\gamma^0 m\tilde{b}$ . (39)

Each of the four Hamiltonion density terms defined in (38), (28) and (39) independently commutes with $\tilde{G}^a(x)$, so each can operate on $|\tilde{0}_\Lambda\rangle$ to make gauge-invariant states "inside" the exponent $\exp(iO_\psi)$.

But in addition to these Hamiltonian density operators, it is possible to define the following local "meson-like", "baryon-like", and "anti-baryon-like" creation operators "inside" the exponent $\exp(iO_\psi)$:

$$\tilde{M}^{0\dagger}_{fg}(\vec{x}) \equiv \tilde{b}^\dagger_{f\alpha s}(\vec{x})\gamma^0\tilde{d}^\dagger_{g\alpha s}(\vec{x}) \qquad \text{(spin 0)}$$

$$\tilde{M}^{1\dagger}_{fg}(\vec{x}) \equiv p^i\tilde{b}^\dagger_{f\alpha s}(\vec{x})\gamma^0\gamma^i\tilde{d}^\dagger_{g\alpha s}(\vec{x}) \qquad \text{(spin 1)}$$

$$\tilde{B}^\dagger_{fghstu}(\vec{x}) \equiv \varepsilon^{\alpha\beta\gamma} \tilde{b}^\dagger_{f\alpha s}(\vec{x}) \tilde{b}^\dagger_{g\beta t}(\vec{x}) \tilde{b}^\dagger_{h\gamma u}(\vec{x})$$

$$\tilde{\bar{B}}^\dagger_{fghstu}(\vec{x}) \equiv \varepsilon^{\alpha\beta\gamma} \tilde{d}^\dagger_{f\alpha s}(\vec{x}) \tilde{d}^\dagger_{g\beta t}(\vec{x}) \tilde{d}^\dagger_{h\gamma u}(\vec{x}). \qquad (40)$$

Here $\alpha, \beta, \gamma$ are fundamental color indices, $f, g, h$ are flavor indices, and $s, t, u$ are 4-spinor indices, all of which were suppressed in the definitions of (30) and (34). For the baryon and anti-baryon operators to be gauge-invariant, the flavor-spin index combination must be symmetric. These local operators then satisfy

$$\left[\tilde{G}(y), \tilde{M}^{n\dagger}_{fg}(\vec{x})\right] = \left[\tilde{G}(y), \tilde{B}^\dagger_{fghstu}(\vec{x})\right] = \left[\tilde{G}(y), \tilde{\bar{B}}^\dagger_{fghstu}(\vec{x})\right] = 0, \qquad (41)$$

so they can be used to build gauge invariant states.

With these operators in hand, one can define a new set of physical states as follows: A physical state is any state which is constructed by using $\exp(iO_\psi)|\tilde{0}_\Lambda\rangle$ or any combination of the operators in (27), (28), (39), and (40) applied to $|\tilde{0}_\Lambda\rangle$ with a factor of $\exp(iO_\psi)$ on the far left. For example, $\exp(iO_\psi)\tilde{M}^{0\dagger}_{fg}(\vec{x})|\tilde{0}_\Lambda\rangle$ is a physical state. Even though this set of physical states may not be the complete set that satisfies the Dirac condition, the set is a consistent set in the sense that the Hamiltonian acting on any physical state produces another physical state.

As an aside, it is interesting that one can create an exact eigenstate (with infinite norm) of the fully interacting QCD Hamiltonian with massless quarks. Since each term of $\tilde{\mathcal{H}}_\psi(\vec{x})$ in (39) has a quark destruction operator, $\tilde{\mathcal{H}}_\psi(\vec{x})$ annihilates the state $|\tilde{0}_\Lambda\rangle$ (and also the state $|\Pi_0\rangle$). Combining this result with the effect seen in (25) of the pure-gauge operator $O_0$, one can see that $\exp(iO_\psi)\exp(O_0)|\Pi_0\rangle$ is an exact eigenstate of the fully interacting QCD Hamilton with massless quarks. The eigenvalue is zero. As mentioned in the pure-gauge QCD section, the problem with this state is that it has infinite norm. If the norm of the state is regulated by the introduction of $\exp(O_\Lambda)$, the state ceases to be an exact eigenstate, but it may still be interesting to explore the expectation value of the energy density of such a state and its dependence on $\Lambda$.

**Conclusions**

A new method has been used to construct exactly gauge invariant QCD states. Two striking features of the method are that it requires the insertion by hand of a mass scale $\Lambda$ and that gauge invariance requires quarks to cluster into meson-like, baryon-like, or anti-baryon-like combinations at every point of space.

One intriguing direction of future study would be to create a gauge-invariant perturbation theory. In normal perturbation theory, the gauge invariant term $-\bar{\psi}\gamma^i i D_i \psi$ in the Hamiltonian is split into a free term $-\bar{\psi}\gamma^i i \partial_i \psi$ and an interacting term $-g\bar{\psi}\gamma^i A_i^a \lambda^a \psi$, neither of which are separately gauge invariant. However, the fact that the state $|0_\Lambda\rangle$ constructed above is an exact eigenstate of $\bar{\psi}\gamma^i i D_i \psi$ introduces the possibility of treating the whole gauge invariant term $\bar{\psi}\gamma^i i D_i \psi$ as the "free" Hamiltonian and the gauge invariant mass term $\bar{\psi} m \psi$ as the interacting Hamiltonian. One advantage of this approach is that gauge invariance would be manifestly maintained at each order of perturbation theory. The small parameter in this approach would be the quark mass divided by a combination of quark momentum and $\Lambda_{QCD}$ (assuming $\Lambda_{QCD}$ sets the scale of the gauge field). Such a gauge invariant perturbation theory could potentially also provide more insight into hadronization processes where standard perturbative QCD starts to break down.

Since the gauge invariant states derived here feature quark clustering no matter how small the coupling constant, this analysis supports the conclusion previously made in [5] that confinement is an unavoidable consequence of gauge invariance and that there should be no deconfinement phase transition at weak coupling. It has been found both experimentally and in lattice calculations that the quark-gluon plasmas (QGPs) created in high energy heavy ion collisions are highly correlated [16-19]. This result is different than one would expect to find for a weakly coupled gas described by a partition function made from single particle states. It would be interesting to use multi-quark states like the ones presented here to build a gauge invariant partition function and see if it would predict a plasma with the kind of strong correlations seen in experiments.